\documentclass[a4paper,12pt]{article}
\usepackage{graphicx}
\usepackage[affil-it]{authblk}
\usepackage[top=1in, bottom=1in, left=1in, right=1in]{geometry}
\usepackage{color}
\usepackage{txfonts}
\usepackage{amssymb}
\usepackage{txfonts}
\usepackage{array}

\begin{document}

\title{The spreading ability of nodes towards localized targets in complex networks}

\author{Ye Sun}
\author{Long Ma}
\author{An Zeng \thanks{anzeng@bnu.edu.cn} }
\author{Wen-Xu Wang}
\affil{
School of Systems Science, Beijing Normal University, Beijing 100875, P.R. China
}

\maketitle
\begin{abstract}
As an important type of dynamics on complex networks, spreading is widely used to model many real processes such as the epidemic contagion and information propagation. One of the most significant research questions in spreading is to rank the spreading ability of nodes in the network. To this end, substantial effort has been made and a variety of effective methods have been proposed. These methods usually define the spreading ability of a node as the number of finally infected nodes given that the spreading is initialized from the node. However, in many real cases such as advertising and medicine science the spreading only aims to cover a specific group of nodes. Therefore, it is necessary to study the spreading ability of nodes towards localized targets in complex networks. In this paper, we propose a reversed local path algorithm for this problem. Simulation results show that our method outperforms the existing methods in identifying the influential nodes with respect to these localized targets. Moreover, the influential spreaders identified by our method can effectively avoid infecting the non-target nodes in the spreading process.
\end{abstract}

\linespread{2}

\newpage
\section{Introduction}
Spreading is a fundamental dynamical process in real systems. It has been intensively studied in many different fields including physics, chemistry, social science, biology and computer science~\cite{Dorogovtsev2008}. The reason behind this is that the emergence of many complex and heterogeneous connectivity patterns in a wide range of biological and social systems can be modeled and investigated by the spreading process in complex networks~\cite{kleineberg2014evolution}. Examples include the epidemic contagion~\cite{keeling2005networks} and rumor/news propagation~\cite{pastor2001epidemic,PhysRevE.69.066130}. After more than a decade of study, our understanding on the properties of spreading processes in complex networks is now much deeper. Results are fruitful. For instance, the spreading on complex networks is found to be a second-order phase transition, and the critical infection rate can be approximated by the mean-field solution based on node degree~\cite{PhysRevLett.105.218701}. The networks with heterogeneous degree distribution in general has a lower critical infection rate than those with homogeneous degree distribution~\cite{moreno2002epidemic}. The spreading records have also been applied to reconstruct the propagation networks~\cite{shen2014reconstructing}. In addition, some methods have been developed to predict the spreading coverage~\cite{colizza2006modeling} and the predictability of the spreading has been discussed~\cite{holme2015time,perez2012prediction}. For a very recent comprehensive review, see ref.~\cite{RevModPhys.87.925}.

Recently, a large amount of attention has been paid to investigate the spreading ability of nodes in complex networks. Identification of the influential spreaders can, for example, help to design a better advertising strategy and a more efficient immunization strategy~\cite{chen2008finding,schneider2011suppressing}. The traditional centrality measures can be naturally applied for this problem. In a pioneer paper~\cite{kitsak2010identification}, the authors pointed out that the k-shell methods can significantly outperform the traditional centralities such as degree~\cite{anderson1992infectious} and betweenness~\cite{friedkin1991theoretical}. After this work, a series of methods have been proposed~\cite{hebert2013global,bauer2012identifying}. For instance, the mixed degree decomposition method consider both the residual degree and the exhausted degree when decomposing the network and rank the nodes accordingly~\cite{zeng2013ranking}; the iterative resource allocation method incorporates the centrality information of neighbors in ranking spreaders~\cite{ren2014iterative}; the path diversity has also been introduced to design the ranking method~\cite{chen2013path}. When spreading starts from multiple origins, the set of nodes with high spreading ability is not easy to find. So far, a number of papers have been devoted to solve this problem~\cite{zhao2014identifying,shuai2012multiple}.

Despite the fact that the existing works on influential spreaders have greatly deepened our understanding of the spreading process in the microscopic level and led to many useful algorithms, one of the key problems still stays untouched, i.e. what would happen if the spreading process does not aim for all the nodes but only suppose to infect a small number of localized target nodes. This is an important research question from both theoretical and practical point of view. In recent literature, the problem of localized targets in many network research has been intensively studied and was found to be very different from the global targets problem~\cite{RevModPhys.87.925}. Examples include the target control of complex networks~\cite{gao2014target} and localized attack on networks~\cite{PhysRevE.65.056109}. Meanwhile, solving this theoretical problem will help us improve the methods for many practical issues. In advertising, for instance, the spreading of the advertisement on beers should cover as much as possible the potential adult customers but avoid propagating to kids. In medical science, also, the spreading of the drug in human body should aim for the ill part but not the healthy part.

In this paper, we investigate the spreading ability of nodes towards localized targets in complex networks. We find that the existing methods for detecting influential spreaders all work poorly in this problem. We thus propose a reversed local path (RLP) algorithm which ranks the spreading ability of nodes by computing the local paths from the target nodes to other nodes. The method is validated with both artificial networks and real networks. The results show that our method can remarkably outperform the existing methods such as degree, k-shell and betweenness in identifying the nodes with high spreading ability towards the localized targets. Moreover, the influential spreaders identified by our method can effectively avoid infecting the non-target nodes in the spreading process. Besides the effectiveness, our method has advantage in the computational complexity compared to the existing methods. Though we consider the classic Susceptible-Infected-Recovered model in this paper, we believe that our method also works well in other spreading models and will have many practical applications in real systems.

\section{Spreading with localized targets}
We first briefly describe the problem of spreading towards localized targets in complex networks. We consider a real network (e.g. the collaboration network of researchers working in network science) as shown in Fig. 1. Two groups of pink nodes are selected as the targeted nodes that we aim to infect. As they are mainly connected with each other, we call them localized targets. Besides these targets, the nodes with the highest degree, betweenness and k-shell values are also highlighted respectively. It is clear that these nodes are topologically far away from the target nodes, the virus or information starting from them has to pass through a lot of non-target nodes to reach the target nodes. If the infection rate is low, the spreading starting from these three nodes may even die out before reaching any of these target nodes. Therefore, the three nodes with highest centralities are no longer the best spreaders towards the localized targets.

We then quantitatively study the difference between the spreading with localized targets (i.e. a small group of nodes are targets) and globalized targets (i.e. all the nodes in the network are targets). To this end, we first define the spreading ability $\rho_i$ of a node $i$ as the fraction of infected target nodes given the spreading originated from node $i$. In this paper, we employ the Susceptible-Infected-Recovered (SIR) model~\cite{Dorogovtsev2008} to simulate the spreading process on networks. The dependence of $\rho_i$ on the spreaders' degree in Barabasi-Albert (BA) networks~\cite{barabasi1999emergence} with the globalized target case and the localized target case is shown in Fig. 2(a)(b), respectively. In Fig. 2(a), i.e. the globalized target case, one can see that $\rho_i$ strongly correlates with the spreaders' degree $k_i$. However, in the localized target case, the correlation between $\rho$ and $k$ is much weaker as shown in Fig. 2(b). For a fixed degree, there is a wide spread of $\rho$ values, which indicates that degree is no longer a good predictor of nodes' spreading ability. In Fig. 2(b), the color of each point represents the mean shortest path length $\langle d_i \rangle$ from the spreader $i$ to the target nodes. One can see that the nodes with small $\langle d_i \rangle$ and large $k_i$ tend to have high $\rho_i$.

To further understand above observations, we investigate the effect of different location of the targets in Fig. 2(c)(d). We fix the number of target nodes as 30 and consider two scenarios, i.e. either the targets are randomly located in the network or they are located in a small area. To realize the second scenario, we first randomly pick up a node and set it as a center for this small area. The rest of the targets are placed in the nodes with the shortest path length not larger than $2$ to the central node. We compare the fraction of infected target nodes $\rho$ as a function of the infection rate $\lambda$ in these two scenarios. As a benchmark, we also plot $\rho$ versus $\lambda$ with the globalized targets in both Fig. 2(c) and (d). One can see that if the 30 targets are distributed randomly, the curve overlaps well with the curve of the globalized target case. However, when the targets are localized within two step distance, the $\rho$ curve is a bit higher than two cases above. These results indicate that the localization of the targets makes the spreading properties significantly differs from the traditional case. In the following, we will mainly focus on how to accurately identify the node with high capability to spread the virus/information to the localized targets.

\section{Methods}
\subsection{Existing methods and extensions}
In spreading dynamics, several centrality indices are widely used to identify the influential spreaders in networks. The basic idea is that the spreading originated from the node with high centrality will finally reach more nodes. In this paper, we consider three representative centrality measures.

(i) \textbf{Degree centrality.} The degree of node $i$ can be defined as $k(i)=\sum\limits_{j\in {G}}a_{ij}$ where $a_{ij}$ is a component of the network's adjacency matrix. Degree represents the number of neighbors this node has, which reflects the direct influence of this node to others.

(ii) \textbf{Betweenness centrality.} The betweenness centrality of node $i$, $b_{i}$, is defined as follows: Between every combination of nodes $a$ and $b$ excluding $i$, we can obtain at least one shortest paths. After respectively defining the number of all these paths and the paths though node $i$ to be $n_{a,b}$ and $n_{a,b}(i)$, $b_{i}$ is then given by:
\begin{equation}
\label{eq:1}
b_{i}=\sum\limits_{(a,b)}\frac{n_{a,b}(i)}{n_{a,b}}
\end{equation}

(iii) \textbf{k-shell decomposition.} By removing nodes with degree less than or equal to $k$ iteratively, the k-shell (also called k-core) method tends to have lower implementation complexity than betweenness and higher accuracy than both degree and betweenness. The definite operations are as follows: We start by removing nodes with degree $ k=1 $ until there is no node left with $ k=1 $ in the network. Then the k-shell value of these removed nodes is set as $ k_{s}=1 $. In step $n$, one should continually remove nodes with residual degree no more than $n$. According to the above operation, the nodes removed in step $ n $ have a k-shell value $ k_{s}=n $.

(iv) \textbf{Local degree.} Considering the findings in Fig. 2 that both degree and distance are essential factors affecting the spreading ability of nodes towards the localized targets, here we consider an additional index based on degree, called local degree (LD) $k_l$. It is simply defined as the degree of the node directly connecting to the target nodes.

\subsection{The reversed local path method}
In order to better identify the spreaders that can easily infect the localized target nodes, we put forward a reversed local path (RLP) method. The basic idea for RLP is to compute the paths up to length $3$ starting from the target nodes to other nodes. The paths with different lengths are aggregated to obtain the final score of a node. The nodes with large final score has high spreading ability towards the target nodes. The method is called reversed local path because only the relatively short paths are taken into account and the paths are counted in the opposite direction to the spreading process (i.e. calculation is from spreaders to target nodes in real spreading, but from target nodes to spreaders in RLP). Mathematically, the formula for RLP reads
\begin{equation}
S_{RLP}=\sum_{l=0}^{2}\epsilon^l fA^{l+1},
\end{equation}
where $f$ is a $1\times N$ vector in which the components corresponding to the target nodes are $1$, and $0$ otherwise. $\epsilon$ is a tunable parameter controlling the weight of the paths with different lengths. In fact, the introduction of parameter $\epsilon$ is inspired by the well-known Katzs index~\cite{katz1953new}. Usually, $\epsilon$ is set to be a small value. In this paper, we fix $\epsilon=0.1$. We only take into account the paths with small length for the sake of efficiency. We have checked that if we extend the path length to $10$, the results will not be much better, sometimes even worse, depending on the setting of $\epsilon$. In fact, the reversed computation (i.e. from target nodes to spreaders) can also significantly reduce the computational complexity, especially when the targets are few and the network is very large. The computational complexity to traverse the neighborhood of a node is simply $k$. If one estimates the spreading ability of each node by directly computing their local paths to target nodes, the computational complexity is $O(Nk^3)$ where $N$ is the number of nodes in the network. However, with RLP the computational complexity can be reduced to $O(mk^3)$ where $m$ is the number of the targets. As $m<<N$ in the localized target problem, the RLP is much more efficient. The RLP process is illustrated with a toy network in Fig. 3. One can see that the most highly ranked node by RLP is different from the nodes with maximum degree and maximum k-shell.

\subsection{Data and Metric}
To validate the RLP method, we will apply it to both artificial and real networks. The artificial networks include the well-known Watts-strogatz (WS) model~\cite{watts1998collective} and Barabasi-Albert (BA) model~\cite{barabasi1999emergence}. We also consider 10 real networks from both social and nonsocial systems. Social networks are: Dolphins (friendship)~\cite{lusseau2003bottlenose}, Jazz (musical collaboration)~\cite{gleiser2003community}, Netsci (collaboration network of network scientists)~\cite{newman2006finding}, Email (communication)~\cite{guimera2003self}, Blog (online blog network of politicians)~\cite{adamic2005political}. Nonsocial networks are: Word (adjacency relation in English text)~\cite{newman2006finding}, E. coli (metabolic)~\cite{jeong2000large}, C. elegans (neural network)~\cite{duch2005community}, TAP (yeast protein-protein binding network generated by tandem affinity purification experiments)~\cite{gavin2002functional}, Y2H (yeast protein-protein binding network generated using yeast two hybridization)~\cite{jeong2001lethality}. Throughout this paper, we present the results of the two artificial networks and two selected real networks (i.e. Netsci and Y2H). The results of the other real networks are reported in Table 1.

For all the methods mentioned above, we generate the final ranking of nodes. In principle, a well-performing ranking algorithm should obtain a ranking as consistent as possible with the ranking based on nodes' spreading ability $\rho$. We then use the Kendall's tau rank correlation coefficient ($\tau$)~\cite{kendall1938new} to estimate how a certain obtained ranking is correlated to the ranking by the true spreading ability $\rho$ of nodes. According to the definition of Kendall's tau coefficient, $-1\leq\tau\leq1$. In the most ideal case where $\tau = 1$, for each pair of two nodes $i$ and $j$, if $i$ is ranked higher than $j$ by the method, the spreading originated from $i$ will cover more targets than the spreading starting from $j$.

\section{Results}
To begin our analysis, we first compare the accuracy $\tau$ of the above-mentioned ranking methods under different infection rate $\lambda$ in Fig. 4. We consider the case where there are 30 randomly distributed targets in the network. Four networks are considered. In WS and BA networks, we do not show the results of the k-core method as the k-shell values of all the nodes in these two networks are almost the same. One immediate observation in Fig. 4 is that the RLP method has much higher accuracy $\tau$ than the other methods, especially when $\lambda$ is small. However, when $\lambda$ is too large and far exceeding the critical infection rate $\lambda_{c}$ (marked by the orange vertical dashed lines in the figure), the spreading originated from each may cover nearly the whole network including the target nodes. In this case, the final spreading coverage can no longer reflect the true spreading ability of nodes. Therefore, the $\tau$ value of RLP is similar to that of the other three methods when $\lambda$ is large.

We then compare the performance of RLP and the local degree (LD) method in Fig. 5. The way we place the target node is the same as Fig. 2(b). We first select a node in the network as the so-called central node. There are $m$ targets in the network and the $m-1$ targets randomly locate in the nodes with maximum distance $L$ (measured by the shortest path length) to the central node. Apparently, when $L$ is infinitely large, these $m$ nodes distribute randomly in the network. The smaller $L$ is, the more localized the targets are. Here, we set the value of the infection rate near the critical infection rate $\lambda_c$ in each network. One can see that the RLP method constantly outperforms the LD method. Interestingly, the difference between $\tau$ of RLP and $\tau$ of LD is more obvious when $m$ and $L$ are smaller, indicating that the more localized the targets are, the larger the advantage of RLP over LD is. Moreover, $\tau$ of both RLP and LD tends to increase with $m$ and $L$. This is because the spreading generally covers more target nodes in this situation, making $\rho$ of more spreaders nonzero. As such, the spreading ability of spreaders becomes more distinguishable (as the computation of $\tau$ depends on nonzero $\rho$). Therefore, $\tau$ of both RLP and LD is higher when $m$ and $L$ are larger. This result also indicates that the spreading ability of spreaders are much more difficult to rank when the targets are localized.

Fig. 4 and Fig. 5 only show the results of $\tau$ in four networks. As mentioned above, we actually applied our method to 10 real networks in this paper. The results of all these real networks are summarized in Table 1. One can see that in all the considered real networks, the RLP method outperforms the rest of other methods. In general, the advantage of RLP over other methods are larger in the real networks with high diameter $D$ such as Netsci and Y2H. In these networks, the localized effect of the target nodes is stronger.

In fact, when we try to infect target nodes, some non-target nodes are infected as well. However, in many real systems the propagation towards the non-target nodes should be avoided. For instance, in advertising the beer company should avoid showing their advertisement to the kids when they try to promote their beer sale by posting advertisement in the online social networks. Similarly, in medicine science the drug for cancer should propagate to the tumour cells but avoid harming the healthy cells. Accordingly, we investigate the ability of the RLP method in avoiding infecting non-target nodes and compare the results with the degree and LD methods. For each method, we pick up the most highly ranked node $i$. Given the spreading initialized from $i$, the fraction of finally infected target nodes and non-target nodes are respectively denoted as $\rho_i$ and $v_i$. In Fig. 6, we show the relation between $\rho$ and $v$ under different infection rates when the three ranking methods are applied to four networks. The faster $\rho$ increases with $v$, the better the method is in avoiding infecting the non-target nodes. Clearly, the RLP method outperforms the degree and LD methods as it can achieve a high $\rho$ with a very small $v$. The advantage of RLP is smaller in BA network. This is because the network has one or several hub nodes (nodes with very large degree) and they are very easy to be infected. Once a hub node is infected, many neighboring non-target nodes will be easily infected. Though some other real networks have hub nodes too, these real networks have some level of community structure (like Fig. 1) such that the network diameter is large and the target nodes can form a local structure that is far away from the hub nodes.

\section{Discussion}
Identification of the influential spreaders is a very important problem from both theoretical and practical point of view. Though a number of methods have been proposed in the literature, the basic assumption for these works is that the spreading aims to infect all the nodes. Inspired by the fact that in many real systems only a small number of nodes in the network are intended to be infected, we put forward a target-oriented spreading problem in this paper. We find that this problem is significantly different from the traditional spreading problem in terms of the influential spreader identification. Specifically, the traditional centrality methods such as degree, betweenness and k-shell are found to be inefficient in finding the spreader that can effectively infect the target nodes. We thus propose a reversed local path method to rank the spreading ability of the nodes towards the target nodes. The simulation results indicate that our method can remarkably outperform the traditional methods, especially when the target nodes are relatively few and strongly localized. The methods are validated in both artificial and real networks. Finally, our method is found to be able to effectively suppress the infection to the non-target nodes.

We believe that this paper proposes a very general research problem and many related issues could be studied in the near future. For instance, to better understand the statistical properties of the target-oriented spreading process, one can systematically investigate the effect of target number and the topological distribution of the targets on the epidemic phase transition and the critical infection rate. Moreover, the method in this paper aims to maximize the coverage of the target nodes, a better method could try to maximize this objective and minimize the coverage of the non-target nodes simultaneously. Our method is based on the local paths, a better method might be designed based on the likelihood maximization approach~\cite{altarelli2014bayesian}. In this way, not only a more accurate method could be developed, some theoretical estimation of the final infected nodes given the spreading originated from different nodes could be obtained. Finally, how to control the spreading process towards the target nodes while the virus or information is already propagation in the networks is also a meaningful research issue. We believe that our work serves as a very good starting point for these problems.

\section*{Acknowledgments}
\noindent
{\bf Authors' contributions. } A.Z. designed the research, Y.S. and L.M. performed the numerical simulations, all the authors analyzed the results and wrote the paper.

\noindent
{\bf Funding statement. } This work is supported by the National Science Foundation of China (No. 11547188) and the Young Scholar Program of Beijing Normal University (2014NT38).

\noindent
{\bf Competing interests. } We declare we have no competing interests.

\noindent
{\bf Data accessibility. } The data of all the real networks used in this paper are publicly accessible.


\clearpage
\begin{figure}
 \center
    \includegraphics[width=16cm]{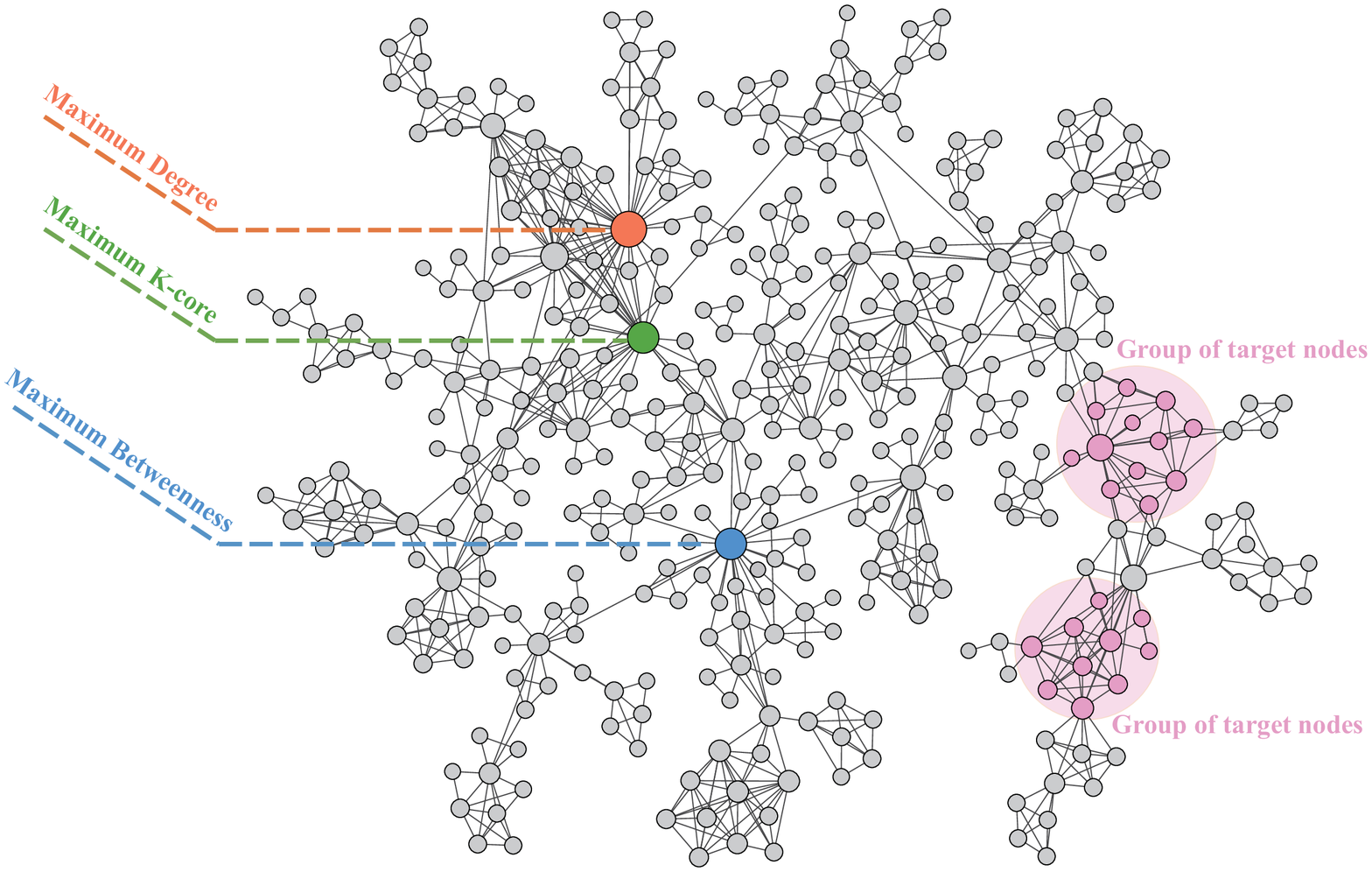}
    \caption{(Color online) Illustration of the problem of spreading towards localized targets in complex networks. The network is the collaboration network of researchers working in network science ($379$ nodes and $914$ links). The pink nodes are the targets that we want to infect. The high centrality nodes are respectively highlighted.}
    \label{Schematical}
\end{figure}

\begin{figure}
 \flushleft
    \includegraphics[width=16cm]{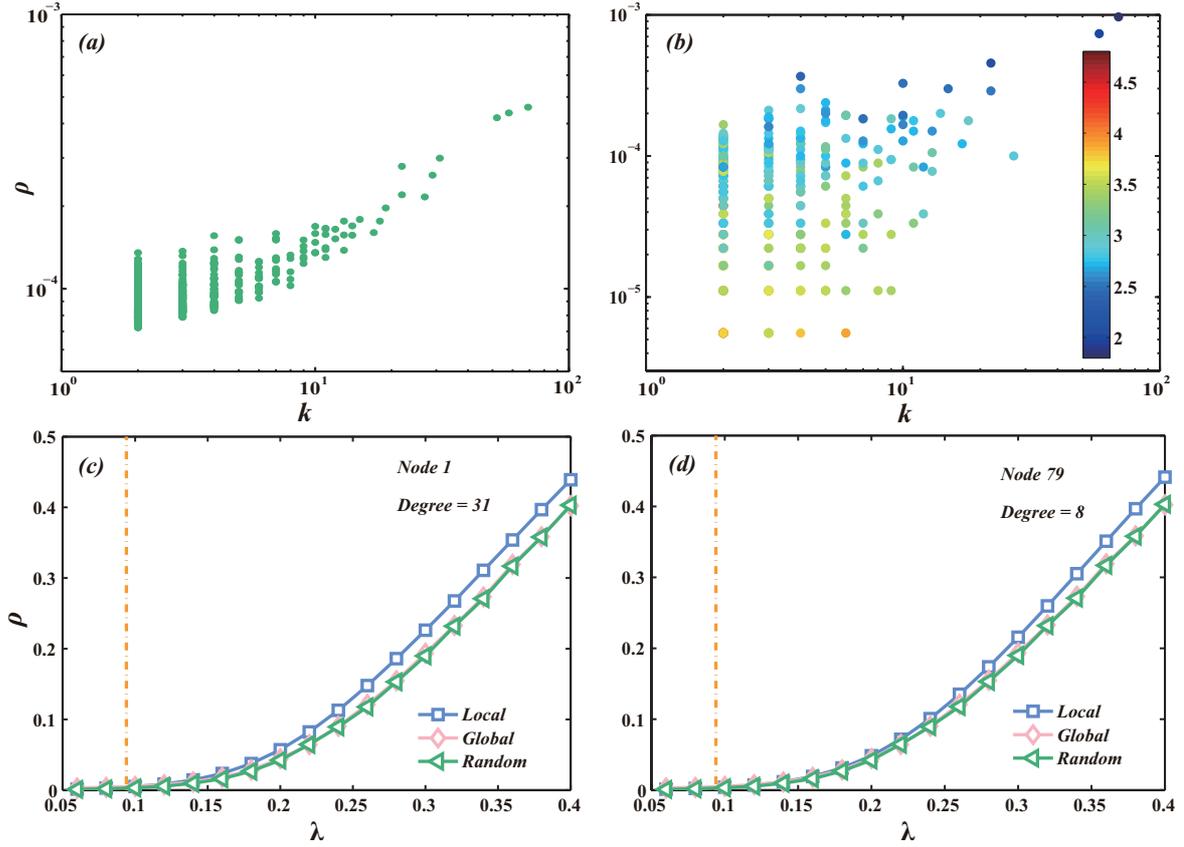}
    \caption{(Color online) (a) The dependence of the fraction of infected target nodes $\rho$ on the initial spreaders' degree $k$. In this sub-figure, all the nodes in the network are target nodes. (b) The dependence of the fraction of infected target nodes $\rho$ on the initial spreaders' degree $k$ and the mean shortest path length $\langle d \rangle$ from the spreader to the target nodes. The color of each point represents the $\langle d \rangle$ of the spreader. In this sub-figure, there are only 30 target nodes. A node is randomly selected as a center and the rest of the targets are placed in the nodes with the shortest path length no larger than 2 to the center. In both (a)(b), the infection rate $\lambda=0.06$, slightly smaller than the critical infection rate $\lambda_c=0.094$. (c)(d) The fraction of infected target nodes $\rho$ as a function of infection rate $\lambda$. In pink rhombus line, all the nodes in the network are target nodes. In green triangle line, we randomly select $30$ nodes as the target nodes, while in blue square line, the target nodes are located the same as (b). The difference between (c) and (d) is that the center has $k=31$ in (c) while $k=8$ in (d). In all sub-figures, the networks are BA model with $N=500$ and $k=4$. The results are obtained by averaging $500$ independent realizations.}
    \label{SanDian}
\end{figure}

\begin{figure}
 \center
   \includegraphics[width=13cm]{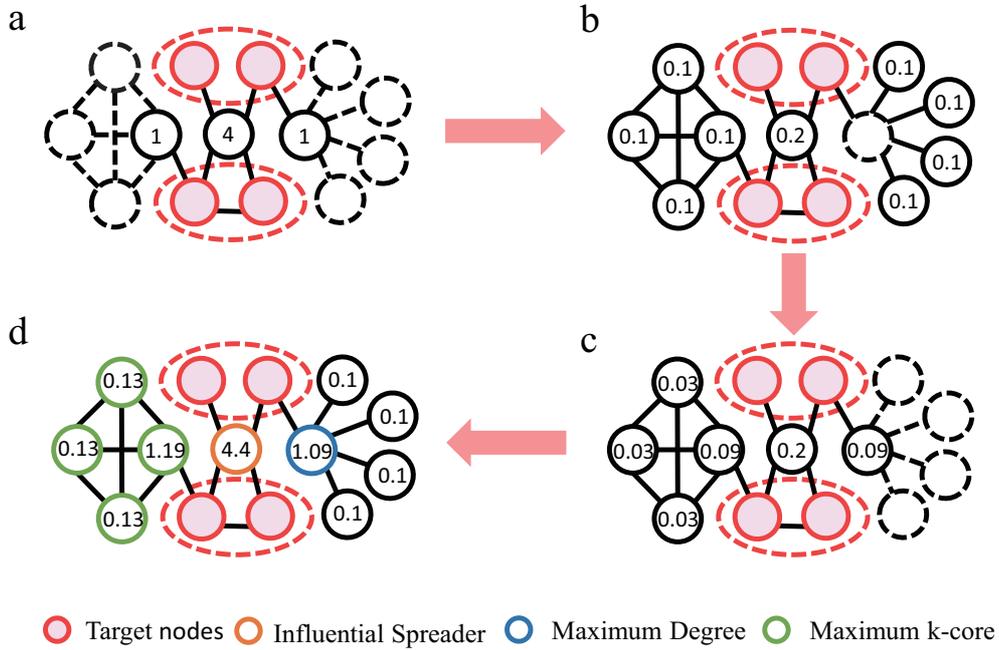}
    \caption{(Color online) Illustrations of the reversed local path algorithm (RLP). The red nodes are target nodes and others are non-target nodes. (a) The nodes with numbers are the first-order neighbors of the targets. All irrelevant nodes and edges are marked in dashed lines. The numbers on the nodes are obtained by computing $fA$. (b) The nodes with numbers are the second-order neighbors of the targets. All possible paths with length 2 are considered and the numbers on the nodes are obtained by computing $\epsilon fA^2$. (c) The nodes with numbers are the third-order neighbors of the targets. All possible paths with length 3 are considered and the numbers on the nodes are obtained by computing $\epsilon^2 fA^3$. (d) The aggregated RLP score of non-target nodes are shown in this figure. The orange, blue and green nodes have maximum RLP, degree and k-core values, respectively.}
    \label{figModel}
\end{figure}

\begin{figure}[t!]
 \center
    \includegraphics[width=16cm]{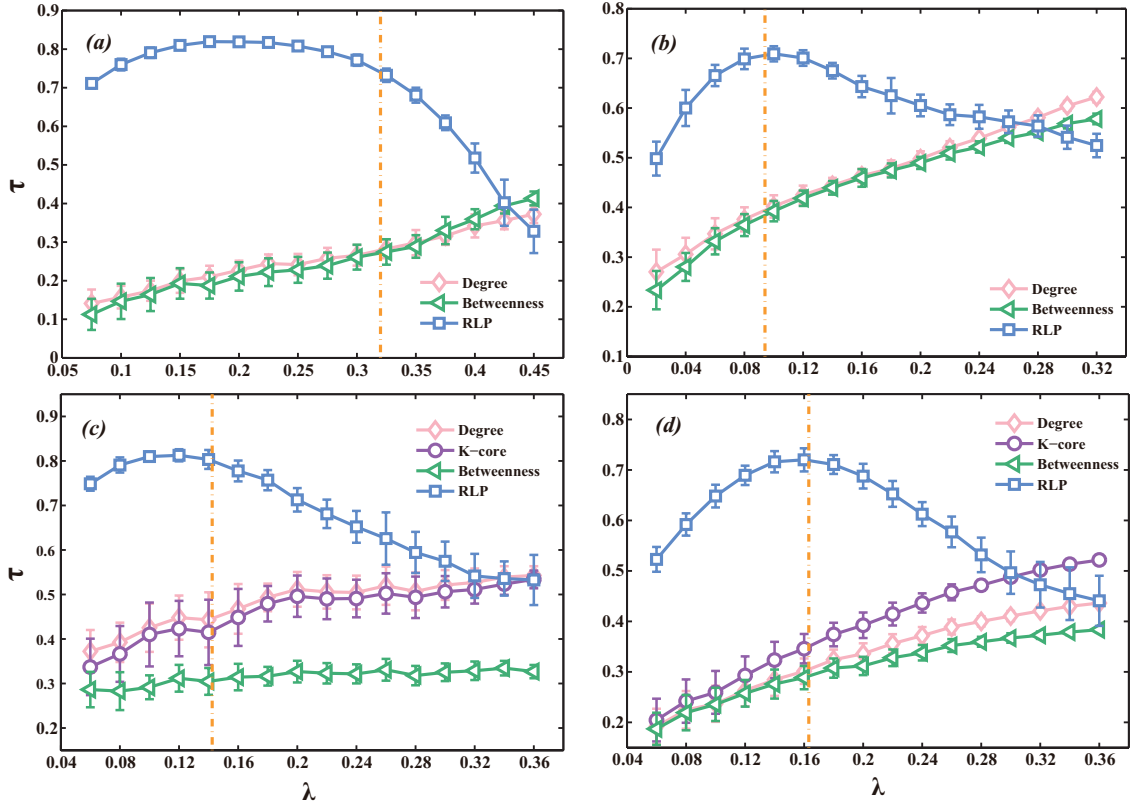}
    \caption{(Color online) Kendall's tau rank correlation coefficient $\tau$ between the rankings obtained from different methods and the true spreading ability $\rho$ under different infection rate $\lambda$. Four networks are considered, i.e. (a) WS, (b) BA, (c) Netsci and (d) Y2H networks. In each network, $30$ target nodes randomly locate in the network. Ranking methods include degree (red diamonds), betweenness (green triangles), k-core (purple circles) or RLP (blue squares) methods. The orange dashed line corresponds to the critical infection rate. The results in each figure is obtained by averaging over $5000$ independent realizations.}
    \label{random}
\end{figure}

\begin{figure}
 \center
    \includegraphics[width=16cm]{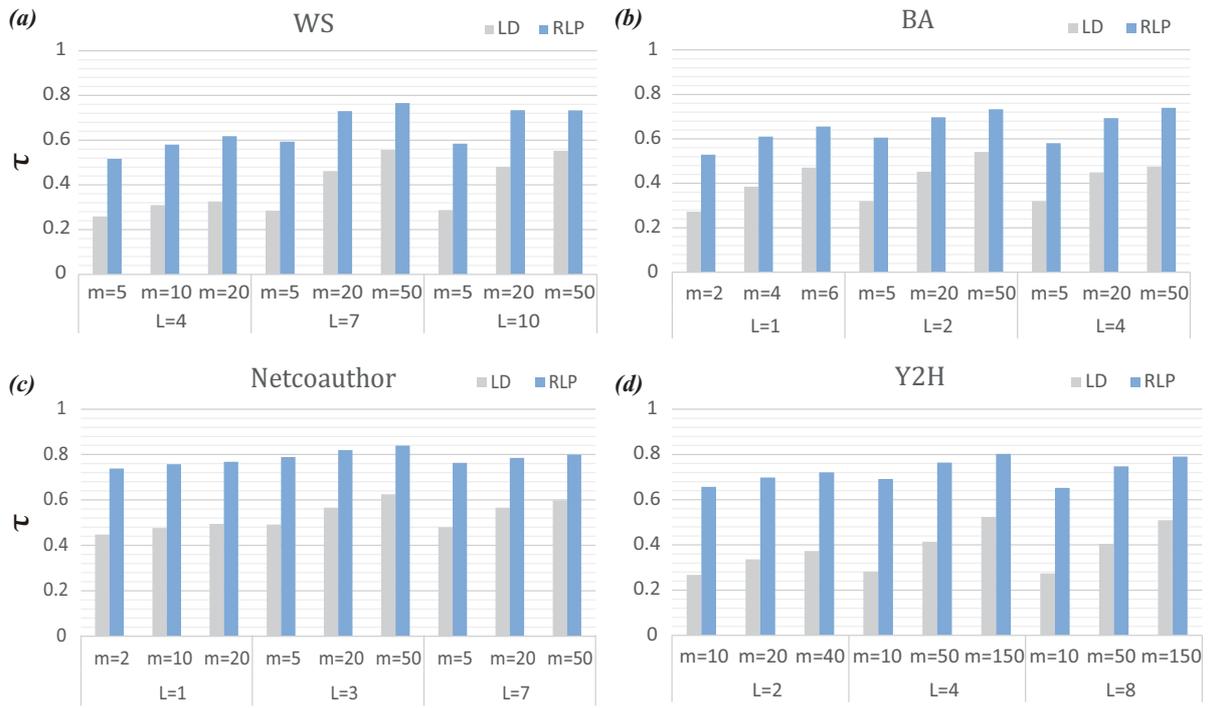}
    \caption{(Color online) The spreading ability ranking accuracy $\tau$ under different $m$ and $L$ in four networks. The parameters for WS and BA networks are $N=500$ and $k=4$. The results in this figure are obtained by averaging over $5000$ independent realizations.}
    \label{zhifangtu}
\end{figure}

\begin{figure}
 \center
    \includegraphics[width=16cm]{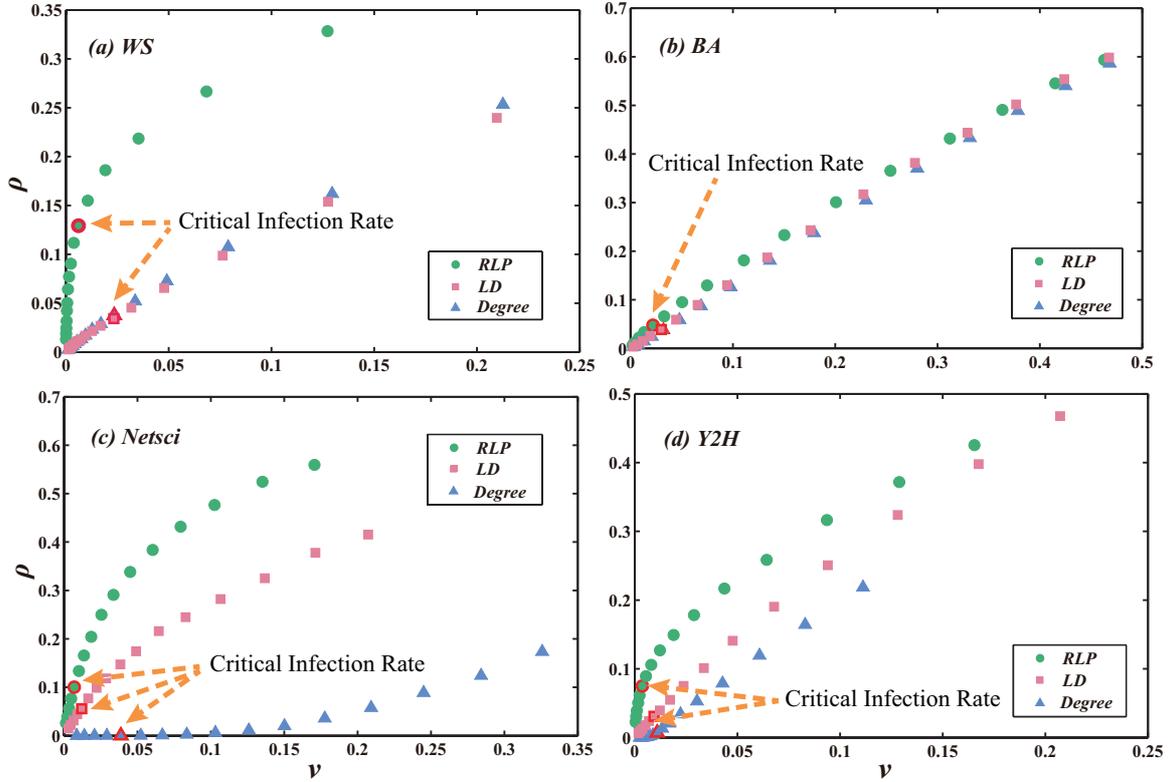}
    \caption{(Color online) The relation between the fraction of infected target nodes $\rho$ and the fraction of infected non-target nodes $v$ under different infection rates when the RLP, LD and degree methods are applied to four networks. Each point in this figure represents the result obtained with a certain infection rate. The point corresponding to the critical infection rate is marked in the figure. In each network, $30$ target nodes randomly locate within distance $L$ to a center node. In WS network, $L=4$ and the center node has degree 4. In BA network, $L=2$ and the central node has degree 7. In Netsci network, $L=2$ and the central node has degree 19. In Y2H network, $L=2$ and the central node has degree $5$. The network parameters for BA and SW are $N=500$ and $k=4$. The results are obtained by averaging over $100$ independent realizations.}
    \label{WuShang}
\end{figure}

\begin{table}[tbp]
\centering
\caption{{\bf Structural properties and ranking results of different methods in ten real networks.} Structural properties include network size ($N$), average degree ($\left \langle k \right \rangle$), network diameter ($D$), critical infection rate $\lambda_{c}$ (which is computed by $\lambda_c=\frac{\langle k\rangle}{\langle k^2\rangle-\langle k\rangle}$). The random scheme represents the case where $10\%$ nodes are set as the target nodes and randomly distributed in the network. The local scheme stands for the case where $10\%$ nodes are set as the target nodes and locate in the nodes with maximum distance $L=2$ (measured by the shortest path length) to a randomly selected central node. According to Fig. 4, we compare $\tau$ of four methods including degree ($\tau_d$), betweenness ($\tau_b$), k-core ($\tau_k$) and RLP ($\tau_{RLP}$) in the random scheme. According to Fig. 5, we compare $\tau$ of two methods including Local degree ($\tau_{LD}$) and RLP ($\tau_{RLP}$). The infection rate for the SIR model in each network is set as $\lambda_c$. The results of the RLP method are highlighted.}
\begin{center}
\renewcommand\arraystretch{0.7}
\begin{tabular}{p{1.7cm}<{\centering}|p{0.8cm}<{\centering}
p{0.8cm}<{\centering}p{0.6cm}<{\centering}p{1cm}<{\centering}|p{1.1cm}<{\centering}p{1cm}<{\centering}p{1cm}<{\centering}p{1.1cm}<{\centering}|
p{1.1cm}<{\centering}p{1cm}<{\centering}}
\hline
\hline
\rule{0pt}{2ex}
   & \multicolumn{4}{c|}{Network properties} & \multicolumn{4}{c|}{Random scheme} & \multicolumn{2}{c}{Local scheme} \\
Network & $N$ & $\left \langle k \right \rangle$ & $D$ & $\lambda_{c}$ & $\left \langle \tau  \right \rangle_{d}$ & $\left \langle \tau \right \rangle_{b}$ & $\left \langle \tau \right \rangle_{k}$ & $\left \langle \tau \right \rangle_{RLP}$& $\left \langle \tau \right \rangle_{LD}$ & $\left \langle \tau \right \rangle_{RLP}$\\[0.6ex]
\hline
\rule{0pt}{0ex}
Dolphins & 62 & 5.13 & 8 & 0.172 & 0.776 & 0.531 & 0.775 & \textbf{0.830} & 0.548 & \textbf{0.757}\\[0ex]
Word & 112 & 7.59 & 5 & 0.078 & 0.764 & 0.639 & 0.754 & \textbf{0.815} & 0.690 & \textbf{0.821}\\[0ex]
Jazz & 198 & 27.70 & 6 & 0.027 & 0.665 & 0.519 & 0.671 & \textbf{0.791} & 0.589 & \textbf{0.835}\\[0ex]
E.coli & 230 & 6.04 & 11 & 0.075 & 0.713 & 0.491 & 0.752 & \textbf{0.840} & 0.572 & \textbf{0.833}\\[0ex]
C.elegans & 297 & 14.46 & 5 & 0.040 & 0.687 & 0.577 & 0.700 & \textbf{0.780} & 0.590 & \textbf{0.780}\\[0ex]
Netsci & 379 & 4.82 & 17 & 0.142 & 0.443 & 0.305 & 0.415 & \textbf{0.803} & 0.575 & \textbf{0.766}\\[0ex]
Email & 1133 & 9.62 & 8 & 0.057 & 0.759 & 0.637 & 0.775 & \textbf{0.799} & 0.622 & \textbf{0.777}\\[0ex]
Blog & 1222 & 27.36 & 8 & 0.013 & 0.708 & 0.607 & 0.713 & \textbf{0.724} & 0.782 & \textbf{0.792}\\[0ex]
TAP & 1373 & 9.95 & 12 & 0.065 & 0.675 & 0.352 & 0.669 & \textbf{0.824} & 0.559 & \textbf{0.733}\\[0ex]
Y2H & 1458 & 2.67 & 19 & 0.163 & 0.301 & 0.289 & 0.346 & \textbf{0.632} & 0.510 & \textbf{0.791}\\[0ex]
\hline
\hline
\end{tabular}
\end{center}
\label{table}
\end{table}

\end{document}